\RequirePackage{lineno}
\documentclass{emulateapj}

\usepackage{amsmath,natbib,graphicx}
\usepackage{epsf}
\usepackage{fancyvrb}
\usepackage{epstopdf}
\usepackage{epsfig}
\usepackage{color}
\usepackage{hyperref}
\DeclareGraphicsExtensions{.jpg,.pdf,.png,.eps,.ps}

\begin{document}


\VerbatimFootnotes
\title{Water Trapping on Tidally Locked Terrestrial Planets Requires Special Conditions}

\shorttitle{Thin-Ice World, Ice Sheet--Ocean World, and Water-Trapped World}

\author{Jun Yang}
\affil{Department of Geophysical Sciences, University of Chicago, Chicago, IL, USA}
\author{Yonggang Liu}
\affil{Woodrow Wilson School of Public and International Affairs, Princeton University, Princeton, New Jersey, USA}
\author{Yongyun Hu}
\affil{Laboratory for Climate and Atmosphere-Ocean Studies, Department of Atmospheric 
 and Oceanic Sciences, School of Physics, Peking University, Beijing, China}
\and
\author{Dorian S. Abbot}
\affil{Department of the Geophysical Sciences, University of Chicago, Chicago, IL, USA}
\email{Correspondence: junyang28@uchicago.edu}

\shortauthors{Yang, Liu, Hu, \& Abbot, ApJL, 2014}

\begin{abstract}
  Surface liquid water is essential for standard planetary
  habitability. Calculations of atmospheric circulation on tidally
  locked planets around M stars suggest that this peculiar orbital
  configuration lends itself to the trapping of large amounts of water
  in kilometers-thick ice on the night side, potentially removing all
  liquid water from the day side where photosynthesis is possible. We
  study this problem using a global climate model including coupled
  atmosphere, ocean, land, and sea-ice components as well as a
  continental ice sheet model driven by the climate model output. For
  a waterworld we find that surface winds transport sea ice toward the
  day side and the ocean carries heat toward the night side. As a
  result, night-side sea ice remains $\cal{O}$(10~m) thick and
  night-side water trapping is insignificant. If a planet has large
  continents on its night side, they can grow ice sheets
  $\cal{O}$(1000~m) thick if the geothermal heat flux is similar to
  Earth's or smaller. Planets with a water complement similar to
  Earth's would therefore experience a large decrease in sea level
  when plate tectonics drives their continents onto the night side,
  but would not experience complete day-side dessication. Only planets
  with a geothermal heat flux lower than Earth's, much of their
  surface covered by continents, and a surface water reservoir
  $\cal{O}$(10\%) of Earth's would be susceptible to complete water
  trapping.
  
\end{abstract}


\section{INTRODUCTION}

Liquid water is necessary for all known life forms
\citep{Kasting2010}.  This has led to the definition of the habitable
zone of a star as the region around it where a planet can maintain
liquid water at its surface. The habitable zone is determined by
radiative calculations \citep{Kastingetal1993, Kastingetal2014} 
assuming the functioning of the silicate-weathering feedback
\citep{Walkeretal1981}. A planet that is in the habitable zone,
however, will not necessarily have surface water because of the
vagaries of water delivery during planet formation
\citep{Morbidellietal2000} and poorly understood processes that
determine the distribution of water between the surface and the mantle
\citep{CowanandAbbot2014}. While too little water is certainly a
problem for life as we know it, too much water is likely problematic
as well. If a planet has so much water that its continents are
submerged, the silicate-weathering feedback cannot function
\citep{Abbotetal2012}, undermining climate stability, and it may be
difficult for life to get started because of a lack of suitable
environments \citep{WardandBrownlee2000}.  An understanding of the
processes that determine the size of the surface liquid water
reservoir is therefore essential for understanding planetary
habitability.

Since M-stars are the most common type of star in the galaxy, it is
important to study issues relevant to the habitability of planets
orbiting them. M-star habitable zone planets may be very dry because
of inefficient scattering of water-bearing planetesimals from beyond
the ice line into the habitable zone during planetary formation
\citep{Lissauer2007, Raymondetal2007}.  Additionally, planets orbiting
small M-stars that start with large eccentricities could suffer
massive heating due to tidal dissipation and lose their water via
atmospheric escape \citep{Barnesetal2013}, and the strong stellar
activity of M-stars may strip the water and atmospheres of habitable
zone planets \citep{Lammeretal2007}.

One issue of particular importance is that due to increased tidal
interactions in closer orbits, planets in the habitable zones of
M-stars will tend to have tidally locked orbital configurations, with
one side always facing the star and the other side in permanent
darkness \citep{Kastingetal1993}. A global atmospheric circulation
with upwelling in the substellar region and downwelling in the rest of
the planet will tend to transport water from the day side to the night
side \citep{Joshietal1997, MerlisandSchneider2010, Edsonetal2011} and
could lead to the trapping of a surface water reservoir on the cold
night side as solid ice \citep{Heathetal1999, Joshi2003}. Recent
calculations using one-dimensional (1D) ice sheet dynamical and
thermodynamical constraints have suggested that a sizeable fraction of
Earth's oceans (equivalent to $\approx$400--900~m of water distributed
globally, or $\approx$15-35\% of Earth's globally distributed surface
water complement of 2700~m) could be sequestered on the night side in
this way \citep{Menou2013}.  These water traps in cold regions may be
so effective that they could persist even if the planet receives more
insolation than the traditional runaway greenhouse limit
\citep{Leconteetal2013}.

As mentioned by \cite{Joshi2003}, \cite{Menou2013}, and
\cite{HuandYang2014}, additional processes including ocean, sea-ice,
and 3D ice-sheet dynamical effects must be considered to fully address
water trapping on the night side of tidally locked planets.  For
example, in the polar regions of Earth, the transport of sea ice into
warmer regions and ocean heat transport underneath sea ice prevent sea
ice from growing thicker than $\approx$5\,m \citep{Lepparanta2005}.
If such processes operate effectively on tidally locked exoplanets, as
the simulations of \cite{HuandYang2014} suggest they can, a shallow
ocean could persist with a thin veneer of sea ice on the night
side. The goal of this paper is to address water trapping using both a
3D global climate model that interactively couples ocean, atmosphere,
sea ice, and land as well as a 3D thermo-mechanical ice sheet model. Our
model results indicate: (1) water trapping is unlikely on planets
without continents if they have at least a few percent of Earth's
water complement and (2) water trapping would only be possible on
planets with continents if those continents cover much of the night
side, the planet has about ten percent of Earth's water complement or
less, and the geothermal heat flux is lower than Earth's.


\section{METHODS}

We perform climate simulations with the Community Climate System Model
version 3.0 \citep[CCSM3;][]{Collinsetal2006}, which was originally developed by
the National Center for Atmospheric Research (NCAR) to study the
climate of Earth. The model contains four coupled components:
atmosphere, ocean, sea ice, and land. The atmosphere component 
calculates atmospheric circulation and parameterizes
sub-grid processes such as convection, precipitation, clouds, and
boundary-layer mixing. The ocean component computes ocean
circulation using the hydrostatic and Boussinesq approximations.  The
sea-ice component predicts ice fraction, ice thickness, ice velocity,
and energy exchanges between the ice and the atmosphere/ocean
\citep{Brieglebetal2004}. The land component calculates surface
temperature, soil water content, and evaporation.

We modify CCSM3 to simulate the climate of habitable planets around
M-stars following \citet{Rosenbloometal2011}, \citet{Liuetal2013}, and
\citet{HuandYang2014}. The stellar spectrum we use is a blackbody with
an effective temperature of 3400~K. We employ planetary parameters
typical of a super-Earth: a radius of 1.5~$R_\oplus$, gravity of
1.38~$g_\oplus$\footnote{\textcolor{black}{We also did three experiments
    with Earth's radius and gravity and different stellar fluxes (866,
    966, and 1066~W~m$^{-2}$). We found that as long as the planet
    does not enter a globally ice-covered snowball state the sea ice
    thickness does not exceed $\cal{O}$(10~m).}}, and an orbital
period of 37~Earth-days. The orbital period of habitable zone planets
around M-stars is roughly $\approx$10--100 days
\citep{Edsonetal2011}. We set the insolation\footnote{This compares to
  $\approx$1370\,W\,m$^{-2}$ for Earth and $\approx$590\,W\,m$^{-2}$
  for Mars.} to 866\,W\,m$^{-2}$ and both the obliquity and
eccentricity to zero. The atmospheric surface pressure is 1.0~bar,
including N$_2$, H$_2$O, and 355 ppmv
CO$_2$\footnote{\textcolor{black}{We tested lower CO$_2$ and stellar
    flux values, for which the model is closer to complete glaciation,
    and found that sea ice is thin in these cases as well.}}. We set
the albedo of sea ice to 0.3 and snow to 0.6, lower than those on
Earth because of the red-shifted stellar spectrum. We set the seawater
freezing point to -1.8\,$^\circ$C.  We use three different continental
configurations: (1) a waterworld (with no continents) and with a
constant ocean depth of 325, 815, or 3800~m; (2) an idealized
supercontinent with a uniform elevation of 100~m above sea level
covering the entire night side and an ocean with a constant depth of
3800~m on the day side; and (3) modern Earth's continental topography
and ocean bathymetry and a substellar point in either the Atlantic
Ocean, Pacific Ocean, \textcolor{black}{or Africa}.

The atmospheric component of CCSM3 has a horizontal resolution of
3.75$^\circ\times$3.75$^\circ$ with 26 vertical levels ranging from
the surface to $\approx$30~km.  The ocean and sea-ice components have
a 3.6$^\circ$ resolution in the West-East direction and a variable
resolution in the South-North direction ranging from 0.9$^\circ$ near
the equator to 4$^\circ$ at high latitudes. The ocean has 25 vertical
layers. We run the model until it reaches statistical steady-state
($\approx$500--2000 Earth years), after which we average all results
for 100 years.


To calculate the thickness of ice sheets over continents, we use the
CCSM3 surface temperature and
precipitation\footnote{\textcolor{black}{We neglect sublimation because
    it is at least an order of magnitude less than precipitation on
    the night side.}}  output to drive a separate ice sheet model. The
ice sheet model we use is the University of Toronto Glacier Systems
Model \citep[UofT GSM;][]{TarasovandPeltier1999,
  TarasovandPeltier2005, LiuandPeltier2013}, which is a 3D
thermo-mechanical model that makes the shallow-ice approximation.  The
ice sheet model calculates ice flow using a standard Glen's flow law
with a temperature-dependent flow factor and includes basal sliding
when the basal temperature is near the pressure melting point. If
basal melting occurs, the meltwater is assumed to be carried by
subglacial streams to the ocean
immediately\footnote{\textcolor{black}{This model does not calculate the
    specific paths meltwater would take as it flows to the ocean;
    however, this is unlikely to affect the equilibrium global ice
    volume.}}.  No surface melting occurs on the night side because
the surface temperature is always less than the melting point.

We use the same planetary parameters in the ice sheet model as in
CCSM3. The spatial resolution of the model is
0.5$^\circ\times$0.5$^\circ$, which corresponds to
80$\times$80~km. The mean density of the bedrock, which is used to
calculate its isostatic deformation, is 3300 kg\,m$^{-3}$ (Earth's
value). We test the influence of the geothermal heat flux by varying
it from 0.01--1.0~W\,m$^{-2}$. The average geothermal heat flux on
continents of modern Earth is 0.07~W\,m$^{-2}$
\citep{DaviesandDavies2010}.  We assume an atmospheric lapse rate of
7.5 K\,km$^{-1}$, and sensitivity tests show that the equilibrium ice
thickness is insensitive to the lapse rate because no surface melting
occurs. Sensitivity tests also show our results are insensitive to
flow and basal sliding enhancement factors.


\section{THIN SEA ICE OVER OCEAN}

\begin{figure*}[]
\begin{center}
\vspace{30mm}
\includegraphics[angle=0, width=160mm]{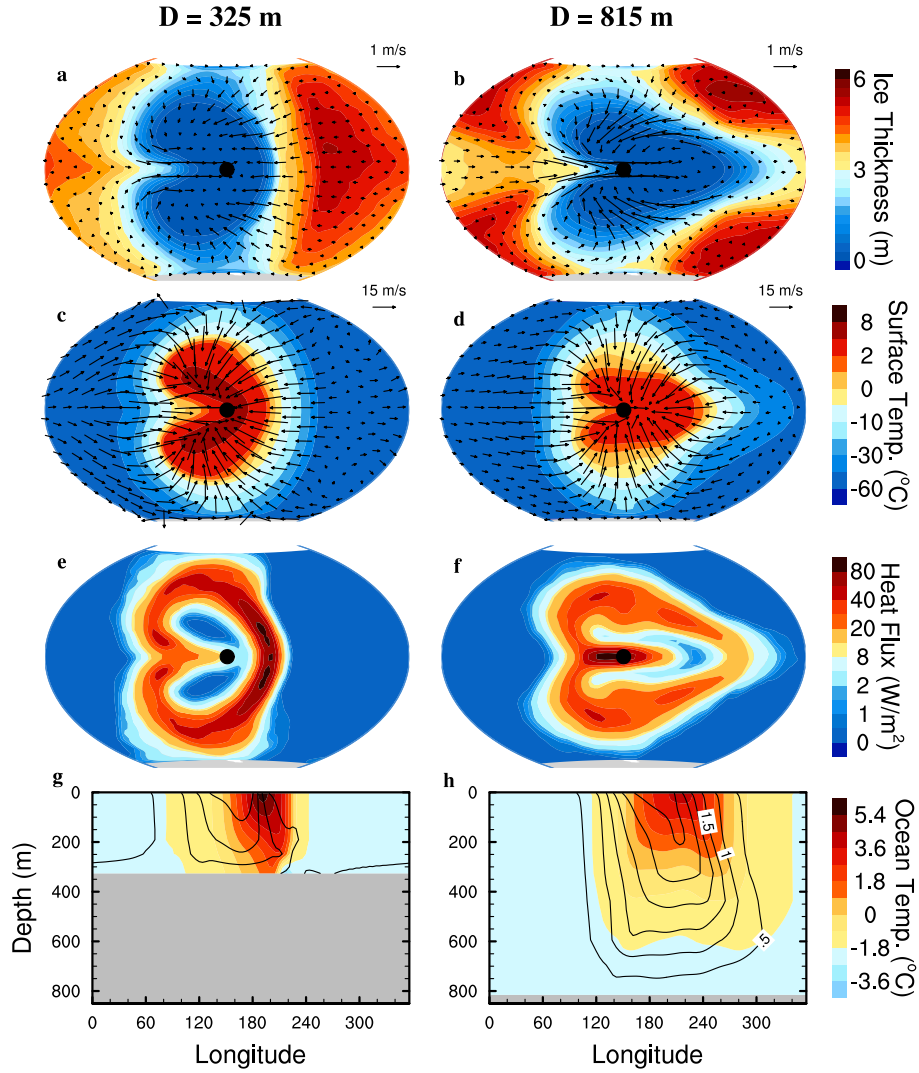}
\vspace{-10mm}
\caption{Climates of tidally locked waterworlds with ocean depths (D)
  of 325~m (left panels) and 815~m (right panels), simulated by CCSM3. 
  (a--b): Sea-ice thickness (color shaded)
  and velocity (vector); (c--d): Surface air temperature (color
  shaded) and surface winds (vector); (e--f): Heat flux from ocean to
  sea ice due to ice bottom melting; and (g-h): Ocean temperature
  (color shaded) and West-East velocity (contours with an interval of
  0.5~m\,s$^{-1}$), averaged between 5$^{\circ}$S and
  5$^{\circ}$N. The black dot in (a--f) is the substellar point. Note
  the color scales for panels (e--h) are not linear. }
\label{fig-sea-ice1}
\end{center}
\end{figure*}

We begin by discussing the sea-ice thickness on a waterworld with a
uniform ocean depth of 325~m, which is below the threshold for which
all of the water would be trapped on the night side if ocean and
sea-ice dynamics are neglected \citep{Menou2013}.  In our calculations
including ocean and sea-ice dynamics, we find a maximum sea-ice
thickness of only 5.4~m (Figure \ref{fig-sea-ice1}(a)), so that very
little of the planetary water complement is sequestered as ice on the
night side\footnote{\textcolor{black}{The model we use cannot calculate ocean dynamics for
``oceans'' as shallow as $\cal{O}$(10~m), but it is possible that
oceans this shallow could be trapped on the night side as ice.}}.

\begin{figure*}[]
\begin{center}
\vspace{20mm}
\includegraphics[angle=0, width=160mm]{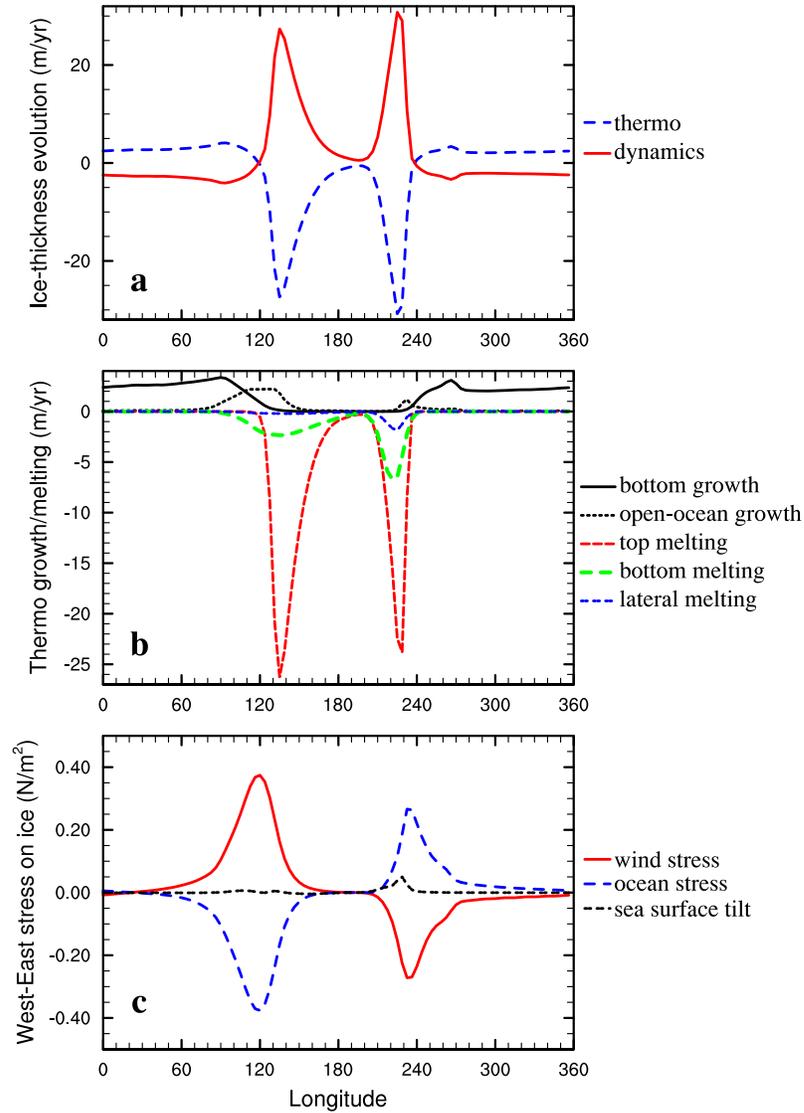}
\vspace{0mm}
\caption{Processes determining sea-ice thickness on a tidally 
locked waterworld with an ocean depth of 325~m. (a) Ice thickness 
evolution due to thermodynamics (blue) and dynamics (red).  
(b) Subcomponents of the thermodynamics: 
bottom ice growth, open-ocean ice growth, top ice melting, 
bottom ice melting, and lateral ice melting. The ice-top snow-to-ice 
conversion term is $\approx$2--3 orders magnitude smaller than other 
terms, so it is not shown. (c) West-East stress on the ice: wind stress, 
ocean stress, and a force due to sea surface tilt. The Coriolis force and 
internal ice stress are very small, so they are not shown. All variables are 
averaged between 5$^{\circ}$S and 5$^{\circ}$N.}
\label{fig-sea-ice2}
\end{center}
\end{figure*}

To understand the thinness of the night-side sea ice, we need to
investigate the dominant balance in the sea-ice time-evolution equation. The
sea-ice thickness is determined by a balance between thermodynamic
growth on the night side and the dynamical export of this ice to the
day side where it is melted (Figure~\ref{fig-sea-ice2}(a)). Ice growth
on the night side is primarily due to freezing at the ice bottom
(Figure~\ref{fig-sea-ice2}(b)). Day-side ice melting occurs both at
the ice top due to warm air (Figure~\ref{fig-sea-ice1}(c)) and at the
ice bottom as a result of warm seawater
(Figure~\ref{fig-sea-ice1}(e)). The simulated ocean circulation of
tidally locked waterworlds is characterized by strong eastward
currents along the equator around the planet
(Figure~\ref{fig-sea-ice1}(g), see also \cite{HuandYang2014}).  This
ocean circulation transports heat from the warm substellar region to
the night side, effectively thinning the ice there, and it becomes
stronger the deeper the ocean (for example, see the 815~m depth case
in Figure~\ref{fig-sea-ice1}(h), and the 3800~m depth case in Figures~2 
and S3 of \cite{HuandYang2014}).  Flow of sea ice from the night side 
to the day side is primarily driven by surface winds
(Figure~\ref{fig-sea-ice2}(c)), which blow toward the relatively warm
substellar region. Sea-ice dynamics are important enough that the
night-side sea ice grows to 200--400~m thick when they are
artificially switched off (not shown). Since sea-ice dynamics can
differ among climate models \citep{Bitz2008} and different
schemes are required for ice hundreds of meters thick
\citep{Tzipermanetal2012}, it is important to check these results in
other models.


To investigate the ability of continents to disrupt ocean and sea-ice
flow, and consequently increase sea-ice thickness, we repeat the
3800~m ocean depth simulation but add continental barrier(s) that are
South-North oriented, narrow, and extend to the bottom of the
ocean. When we add one barrier spanning the western terminator, the
night-side sea-ice thickness increases only slightly (not shown).
When we add an additional barrier on the eastern terminator, so that
both ocean heat transport and sea-ice flow between the night and day
sides are completely blocked, the sea-ice thickness reaches
$\cal{O}$(1000~m). This confirms that ocean and sea-ice dynamics are
essential for maintaining thin sea ice and preventing water trapping
on a waterworld.  These simulations also demonstrate that water
trapping of a significant ocean would be possible if a planet had
continents positioned so that they completely disrupted ocean heat
transport and sea-ice flow between its day and night sides.

\begin{figure*}[]
\begin{center}
\includegraphics[angle=0, width=160mm]{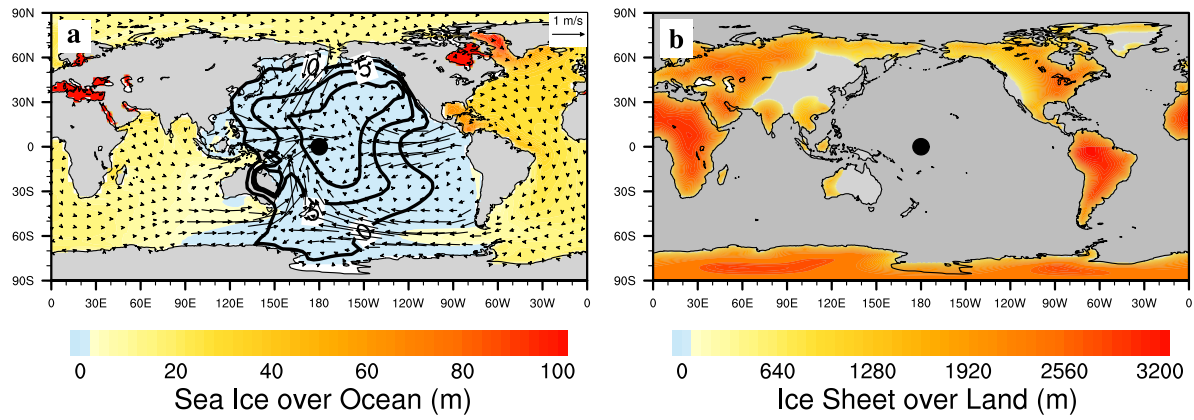}
\vspace{-45mm}
\caption{Thicknesses of sea ice over ocean (a) and ice sheet over land
  (b) for a tidally locked planet with modern Earth's continental
  configuration. In (a), vectors show the sea-ice velocity, and thick black 
  contours indicate surface air temperatures of 0, 5, and
  7\,$^{\circ}$C. In (b), the geothermal heat flux over continents is
  set to 0.05~W\,m$^{-2}$. If it were higher, the ice sheet would be 
  much thinner and flatter. The black dot is the substellar point.}
\label{fig-earth-land}
\end{center}
\end{figure*}

On modern Earth, continents provide large South-North barriers, but do
not completely obstruct ocean and sea-ice flow. Since this may
represent a more realistic possible continental configuration than
complete barriers, we repeat the simulations using modern Earth's
continental configuration. In general, sea ice stays $\cal{O}$(10~m)
thick, although it can grow to $\cal{O}$(100~m) thick
in small regions that are strongly isolated by continents, such as the
Caribbean Sea, Baffin Bay, and the Mediterranean Sea
(Figure~\ref{fig-earth-land}(a)).  Even small gaps in continental
coverage, such as the Drake Passage between South America and
Antarctica and the Bering Strait between Asia and North America, allow
efficient transport of sea ice into warm regions where it can melt.


\section{ICE SHEET THICKNESS OVER CONTINENT} 

\begin{figure*}[]
\begin{center}
\vspace{30mm}
\includegraphics[angle=0, width=145mm]{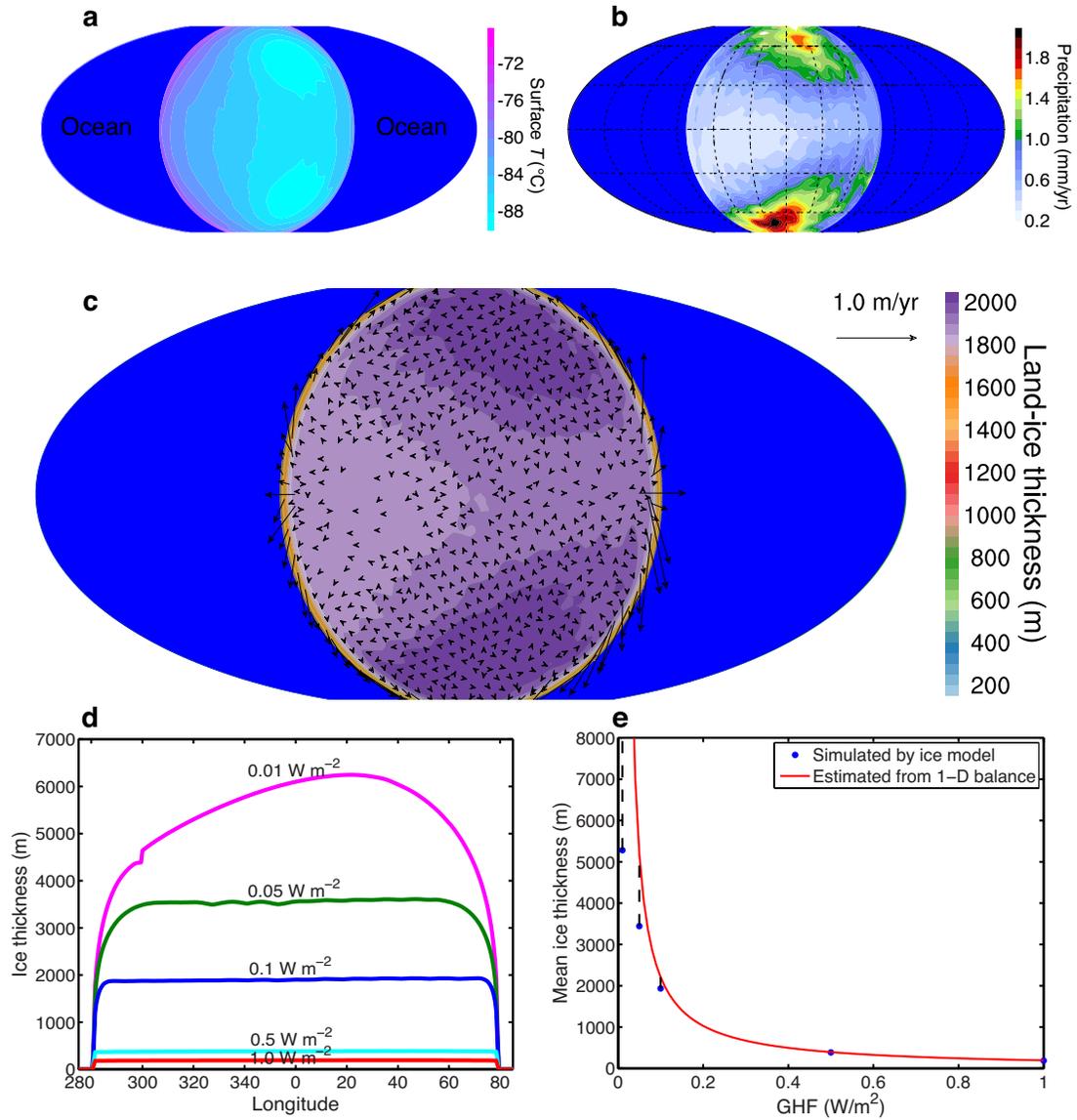}
\vspace{5mm}
\caption{Climate and ice sheet over a supercontinent spanning the
  entire night side. (a) Surface temperature and (b) precipitation
  over the continent, which are both used to drive the ice sheet
  model. (c) Ice sheet thickness (color shaded) and ice sheet flow
  velocity (vector) with a geothermal heat flux (GHF) of 0.1~W\,m$^{-2}$.
  (d) Ice sheet thickness along the equator with GHF ranging from 
  0.01 to 1.0~W\,m$^{-2}$.  (e) Night-side mean ice
  sheet thickness calculated by 1D energy balance limit (red line) and
  by the 3D ice sheet model (dots). The vertical dashed lines in (e)
  indicate the overestimate of ice sheet thickness by the 1D balance
  limit. }
\label{fig-super}
\end{center}
\end{figure*}

We begin our investigation of ice sheets on night-side continents with
the idealized case of a supercontinent that spans the entire night
side, which is optimal for water trapping. Our simulations confirm
that the ice volume trapped on the continent is critically dependent
on the magnitude of the geothermal heat flux, as suggested in
\cite{Menou2013} and \cite{Leconteetal2013}. The average ice
thickness\footnote{The ice sheets on Greenland and Antarctica on
  modern Earth are generally a few kilometers thick. Sea level would
  rise by $\approx$70 m if they were to melt completely
  \citep{Marshall2012}.} on the night side is 3445 m if the geothermal
heat flux is 0.05 W\,m$^{-2}$, and decreases to 387 m if the
geothermal heat flux is 0.5 W\,m$^{-2}$ (Figure~\ref{fig-super}). For
a geothermal heat flux similar to Earth's, a day-side ocean of
$\approx$2880~m (or $\approx$1440~m distributed globally, about half
of Earth's water complement) could be trapped on the night-side
supercontinent.  This estimate agrees fairly well with that obtained
by \cite{Menou2013} using simpler ice sheet models that have the
benefit of elucidating the scaling of ice sheet thickness with a
variety of physical parameters that will vary among exoplanets. If the
geothermal heat flux is significantly higher (as one might expect on a
super-Earth), only a small ocean a few hundred meters deep could be
trapped even on a continent encompassing the entire night side.


The mean ice sheet thickness is well-approximated by the 1D energy
balance limit \citep[e.g.,][]{AbbotandSwitzer2011} if the ice sheet is
thinner than $\approx$2000~m (Figure~\ref{fig-super}(e)). This is
because ice flow is slow and is concentrated around the ice-sheet
edges (Figure~\ref{fig-super}(c)), so it does not have a significant
effect on the equilibrium ice thickness. When the ice sheet is thicker
than $\approx$2000~m, the ice flow speed increases and large regions
of the interior of the ice sheet are affected by ice flow. This
can be inferred from the surface slope of the ice sheet (Figure
\ref{fig-super}(d)) based on the fact that the ice speed is roughly
proportional to the cube of the surface slope. As a result, the
increased export of land ice toward the ocean prevents the ice sheet
from growing to the thickness estimated based on the 1D limit
(Figure~\ref{fig-super}(e)).

When we use modern Earth's continental configuration, a geothermal
heat flux of 0.05 W\,m$^{-2}$, and the substellar point located in the
central Pacific Ocean, the globally distributed ice-sheet thickness is
$\approx$470~m (Figure~\ref{fig-earth-land}(b)). This is 32\% of the
water trapped by the idealized supercontinent spanning the entire
night side (for reference, 36\% of the night side is covered by
continents in this simulation), or 18\% of modern Earth's water
complement.  Moving the substellar point from the Pacific Ocean to the
Atlantic Ocean \textcolor{black}{or Africa} yields lower global-mean ice
sheet thicknesses and similar global-mean sea-ice thicknesses (not
shown).




\section{CONCLUSION} 

\begin{figure*}[]
\begin{center}
\vspace{50mm}
\includegraphics[angle=0, width=160mm]{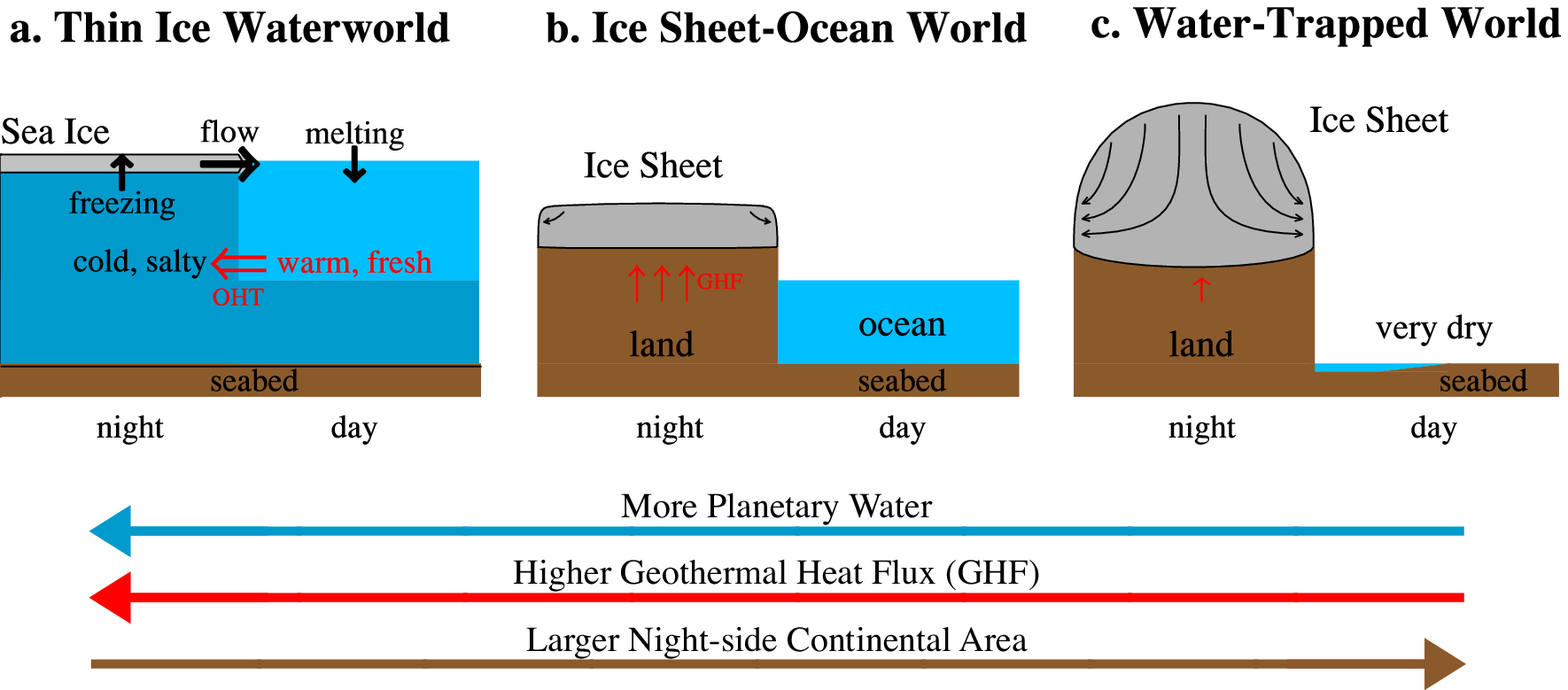}
\vspace{5mm}
\caption{Schematic diagrams of: (a) a planet with thin sea ice on the
  night side of a waterworld, (b) a planet with a small ice sheet on a
  night-side continent, and (c) a water-trapped world with a large
  night-side ice sheet.  In (a) sea-ice dynamics and ocean heat
  transport (OHT) keep the sea ice thin. In (b) the geothermal heat
  flux (GHF) is large enough to keep the ice sheet small so that it
  cannot trap all of the planet's water.  In (c) GHF 
  is small enough that almost all of the water is trapped on the
  night-side ice sheet, although subglacial streams and the melting of
  ice that flows across the terminator allow a small amount of water
  on the day side. This last configuration will only be possible if
  the geothermal heat flux is small, the planetary water complement is
  small, and there are relatively large continents located preferentially on
  the night side.}
\label{fig-schematic}
\end{center}
\end{figure*}

We have applied sophisticated global climate and ice sheet models to
the question of water trapping on the night side of tidally locked
terrestrial planets. We found that ocean heat transport and sea-ice
dynamics are likely to limit night-side sea-ice thickness and prevent
water trapping on a planet with no continents (shown schematically in
Figure~\ref{fig-schematic}(a)).  If a planet has continents that
completely block ocean flow and sea-ice transport between the day and night sides, the
sea-ice thickness can reach the kilometer scale, but even small gaps
in this blocking keep sea ice thin. Additionally, we found that ice
sheets on night-side continents can approach the kilometer scale
(Figure~\ref{fig-schematic}(b)), although this would only limit the
habitability of planets with surface water complements $\cal{O}$(10\%)
of Earth's, very large continents covering most of their night side,
and \textcolor{black}{a geothermal heat flux smaller than Earth's}
(Figure~\ref{fig-schematic}(c)). Such planets would experience
irregular moistening and desiccation of their day side as plate
tectonics exchanges their continents between their day and night
sides. \textcolor{black}{It will be difficult to remotely detect the
  quantity of water trapped on the night sides of exoplanets, but it
  may be possible to determine whether the day sides are dry from the
  bond albedo or by studying the visible/near-infrared phase curves
  \citep[e.g.,][]{Cowanetal2009}, although clouds would complicate such
  observations \citep[e.g.,][]{Yangetal2013}.}

\acknowledgments \textbf{Acknowledgments:} We thank Dawei Li, Raymond
T. Pierrehumbert, Jiping Liu, and Sergey Malyshev for helpful discussions. We thank
Daniel Koll and Douglas MacAyeal for comments on an early version of
this letter. YH is supported by the National Natural Science
Foundation of China (41025018 and 41375072). DSA acknowledges support
from an Alfred P. Sloan Research Fellowship.



\begin{thebibliography}{}
\bibitem[Abbot et al.(2012)]{Abbotetal2012} Abbot, D. S., Cowan, N. B., \& Ciesla, F. J. 2012, \apj, 756, 178
\bibitem[Abbot \& Switzer(2011)]{AbbotandSwitzer2011} Abbot, D. S., \& Switzer, E. R. 2011, ApJL, 735, L27
\bibitem[Barnes et al.(2013)]{Barnesetal2013} Barnes, R., Mullins, K., Goldblatt, C., et al. 2013, Astrobiology, 12, 225
\bibitem[Bitz(2008)]{Bitz2008} Bitz, C. M. 2008, Numerical modeling of sea ice in the climate system, \url{http://www.atmos.washington.edu/~bitz/Bitz_draftchapter.pdf} 
\bibitem[Briegleb et al.(2004)]{Brieglebetal2004} Briegleb, B. P., Bitz, C. M., Hunke, E. C., et al. 2004, 
Technical Note, Document NCAR/TN--463+STR  (Boulder, CO: NCAR)
\bibitem[Collins et~al.(2006)]{Collinsetal2006} Collins, W. D., Bitz, C. M., Blackmon, M. L., et al. 2006, J. Climate, 19, 2122
\bibitem[Cowan et~al.(2009)]{Cowanetal2009} Cowan, N. B., Agol, E., Meadows, V. S., et al. 2009, ApJ, 700, 915
\bibitem[Cowan \& Abbot(2014)]{CowanandAbbot2014} Cowan, N. B., \& Abbot, D. S. 2014, \apj, 781, 1
\bibitem[Davies \& Davies(2010)]{DaviesandDavies2010} Davies, J. H., \& Davies, D. R. 2010, Solid Earth, 1, 5
\bibitem[Edson et al.(2011)]{Edsonetal2011} Edson, A., Lee, S., Bannon, P., Kasting, J. F., \& Pollard, D. 2011, Icarus, 212, 1
\bibitem[Heath et al.(1999)]{Heathetal1999} Heath, M. J., Doyle, L. R., Joshi, M. M., \& Haberle, R. M. 1999, 
Origins Of Life And Evolution Of The Biosphere, 29, 405
\bibitem[Hu \& Yang(2014)]{HuandYang2014} Hu, Y., \& Yang, J. 2014, PNAS, 111, 629   
\bibitem[Joshi et al.(1997)]{Joshietal1997} Joshi, M. M., Haberle, R. M., \& Reynolds, R. T. 1997, Icarus, 129, 450 
\bibitem[Joshi(2003)]{Joshi2003} Joshi, M. M. 2003, Astrobiology, 3, 415
\bibitem[Kasting et al.(1993)]{Kastingetal1993} Kasting, J. F., Whitmire, D. P., \& Reynolds, R. T. 1993, Icarus, 101, 108
\bibitem[Kasting(2010)]{Kasting2010} Kasting, J. F. 2010, How to find a habitable planet (New Jersey, USA: Princeton University Press)
\bibitem[Kasting et al.(2014)]{Kastingetal2014} Kasting, J. F., Kopparapu, R., Ramirez, R. M., \& Harman, C. E. 2014, PNAS, in press
\bibitem[Lammer et al.(2007)]{Lammeretal2007} Lammer, H., Lichtenegger, H. I., Kulikov, Y. N., et al. 2007, Astrobiology, 7, 185
\bibitem[Leconte et al.(2013)]{Leconteetal2013} Leconte, J., Forget, F., Charnay, B., et al. 2013, A\&A, 554, A69
\bibitem[Lepparanta(2005)]{Lepparanta2005} Lepparanta, M. 2005, The drift of sea ice (Heidelberg, Germany:  Springer-Praxis)	                     
\bibitem[Lissauer(2007)]{Lissauer2007} Lissauer, J. J. 2007, ApJL, 660, L149
\bibitem[Liu et al.(2013)]{Liuetal2013} Liu, Y., Peltier, W. R., Yang, J., \&  Vettoretti, G. 2013, Clim. Past, 9, 2579 
\bibitem[Liu \& Peltier(2013)]{LiuandPeltier2013} Liu, Y., \& Peltier, W. R. 2013, J. Geophys. Res., 118, 4410
\bibitem[Marshall(2012)]{Marshall2012} Marshall, S. J. 2012, The Cryosphere (New Jersey, USA: Princeton University Press)
\bibitem[Menou(2013)]{Menou2013} Menou, K. 2013, \apj, 774, 51      
\bibitem[Merlis \& Schneider(2010)]{MerlisandSchneider2010} Merlis, T. M., \& Schneider, T. 2010, JAMES, 2, 13
\bibitem[Morbidelli et al.(2000)]{Morbidellietal2000} Morbidelli, A., Chambers, J., \& Lunine, J. I., et al. 2000, 
Meteoritics and Planetary Science, 35, 1309
\bibitem[Raymond et al.(2007)]{Raymondetal2007} Raymond, S. N., Scalo, J., \& Meadows, V. S. 2007, \apj, 609, 606
\bibitem[Rosenbloom et al.(2011)]{Rosenbloometal2011} Rosenbloom, N., Shields, C., Brady, E., Levis, S., \& Yeager, S. 2011, Technical Note, Document NCAR/TN--483+STR (Boudler, CO: NCAR)
\bibitem[Tarasov \&  Peltier(1999)]{TarasovandPeltier1999} Tarasov, L., \& Peltier, W. R. 1999, J. Geophys. Res., 104, 9517        
\bibitem[Tarasov \&  Peltier(2005)]{TarasovandPeltier2005} Tarasov, L., \& Peltier, W. R. 2005, Nature, 435, 662
\bibitem[Tziperman et al.(2012)]{Tzipermanetal2012} Tziperman, E., Abbot, D. S., \& Ashkenazy, Y., et al. 2012, J. Geophys. Res., 117, C05016
\bibitem[Walker et al.(1981)]{Walkeretal1981} Walker, J. C. G., Hays, P. B., \& Kasting, J. F. 1981, \jgr , 86, 9776   
\bibitem[Ward and Brownlee(2000)]{WardandBrownlee2000} Ward, P. D., \& Brownlee, D. 2000, 
Rare {Earth}: Why complex life is uncommon in the Universe (New York, USA: Copernicus Books)
\bibitem[Yang et al.(2013)]{Yangetal2013} Yang, J., Cowan, N. B., \& Abbot, D. S. 2013, ApJL, 771, L45
\end{thebibliography}
\end{document}